\documentclass[pra,twocolumn,superscriptaddress,showpacs]{revtex4}
\newcommand{\UQ}{School of Mathematics and Physics, University of Queensland, QLD 4072, Australia.}
\newcommand{\UCL}{Department of Physics and Astronomy, University College London, Gower Street, London WC1E 6BT, United Kingdom.}
\usepackage{graphics}
\usepackage{graphicx}    
\usepackage{bm}
\usepackage{amssymb}
\usepackage{bm}
\usepackage{color}
\usepackage{epstopdf}
\usepackage{amsfonts}
\usepackage{amsmath}
\usepackage{amsbsy} 

\newcommand{\etal}{{\em et al.}}

\newcommand{\jpa}{J. Phys. A }

\begin{document}
\title{Spectral Analysis of a Four Mode Cluster State}

\author{S.~L.~W. Midgley}
\affiliation{\UQ}
\affiliation{\UCL}
\author{M.~K. Olsen}
\affiliation{\UQ}

\date{\today}

\begin{abstract}

We theoretically evaluate the squeezed joint operators produced in a single optical parametric oscillator which generates quadripartite entangled outputs, as demonstrated experimentally by Pysher \etal~\cite{pysher}[Phys. Rev. Lett. 107, 030505 (2011)]. Using a linearized fluctuation analysis we calculate the squeezing of the joint quadrature operators below threshold for a range of local oscillator phases and frequencies. These results add to the existing theoretical understanding of this potentially important system.
 \end{abstract}

\pacs{42.50.Dv,42.65.Yj,42.50.Ex,03.67.Lx}

\maketitle

\section{Introduction}

Cluster states are highly entangled multipartite states that have attracted much interest since they were proposed by Raussendorf and Briegel~\cite{raussen_briegel_1} as a resource state with potential applications for one-way quantum computing. However, as with conventional approaches, issues of scalability arise~\cite{Blythe,Walther}. Recent experimental advances have been made by 
Pysher \etal~\cite{pysher} in the generation of a continuous-variable (CV) quadripartite cluster state from a single optical parametric oscillator (OPO), based on a proposal by Menicucci \etal~\cite{men2}. In this approach quantum registers are encoded in the quadratures of the optical frequency comb produced by the spectrum of the OPO. The experimental scheme was shown to be scalable and capable of simultaneously generating 15 quadripartite entangled cluster states. In previous work with others~\cite{nossoquatro,midgley}, we verified the presence of quadripartite entanglement in such a system by demonstrating violations of the van Loock-Furusawa (VLF) criteria~\cite{furu}. We also calculated the squeezed joint quadrature operators in the temporal domain to determine whether the state produced could be classified as a cluster state.

In this Brief Report, we investigate the squeezed joint quadrature operators in the frequency domain as this is the form more amenable to comparison with experiment. A spectral analysis of the system is required as the temporal domain solutions give the squeezing integrated over all frequencies and are not useful when we wish to know for which frequencies the system will be most useful.  By verifying that the squeezed joint quadrature operators approach very low noise levels at particular frequencies, we determine that the scheme approaches an ideal cluster state. This Brief Report is structured as follows. Section II provides details of the theoretical model and method. Section III sets out the results and discussion. A summary of our findings is given in Sec. IV.

\section{SYSTEM AND METHOD}
\label{sys}

The system considered here is a pumped optical cavity containing a $\chi^{(2)}$ non-linear crystal, as investigated theoretically in Ref.~\cite{midgley}. We extend our analysis to incorporate the results of a full spectral analysis in the frequency domain. The optical cavity is pumped by two field modes which give rise to four low-frequency output modes at frequencies $\omega_{3},\omega_{4},\omega_{5},\omega_{6}$. Mode 1 is pumped at a particular frequency and polarization such that it produces modes 3 and 6, as well as modes 4 and 5. Mode 2 is pumped such that it gives rise to modes 5 and 6. 

The master equation for this system is~\cite{Danbook}
\begin{equation}
\frac{\partial \hat{\rho}}{\partial t} = -\frac{i}{\hbar}\Big[\hat{H}_{int}+\hat{H}_{pump}, \hat{\rho}\Big]+\sum_{i=1}^{6}\gamma_{i} {\cal{D}}_{i}[\hat{\rho}]
\label{master}
\end{equation}
where the $\hat{H}_{int}$ is the interaction Hamiltonian given by
\begin{eqnarray}
\hat{H}_{int} = i\hbar[\chi_{1}\hat{a}_{1}\hat{a}_{4}^{\dagger}\hat{a}_{5}^{\dagger} + \chi_{1}\hat{a}_{1}\hat{a}_{3}^{\dagger}\hat{a}_{6}^{\dagger} +\chi_{2}\hat{a}_{2}\hat{a}_{5}^{\dagger}\hat{a}_{6}^{\dagger}]+ \textnormal{h.c.},
\label{Hint}
\end{eqnarray}
with the $\chi_{i}$ representing the effective nonlinearities and $\hat{a}_{i}$ and $\hat{a}^{\dagger}_{i}$ denoting the bosonic annihilation and creation operators for the intra-cavity modes at frequencies $\omega_{i}$. We note here that we are using the full interaction Hamiltonian rather than making a classical undepleted pump approximation leading to linear Heisenberg equations as used  by Pysher \etal~\cite{pysher} (see also the Supplemental Material), which has the advantage that our formalism may also be used above the oscillation threshhold~\cite{nossoquatro,midgley}.  
In a rotating frame, the pumping Hamiltonian, $\hat{H}_{pump}$ is given by
\begin{equation}
\hat{H}_{pump} = i\hbar\sum_{i=1}^{2}\left[\epsilon_{i}\hat{a}_{i}^{\dag}-\epsilon_{i}^{\ast}\hat{a}_{i}\right],
\label{eq:Hpump}
\end{equation}
where $\epsilon_{i}$ are the classical laser amplitudes, the $\gamma_{i}$ represent the cavity losses at each frequency and ${\cal D}_{i}[\hat{\rho}]$ is the Lindblad superoperator~\cite{Danbook} describing zero-temperature Markovian damping. 

\begin{figure}[tbhp]
\includegraphics[width=0.15\textwidth]{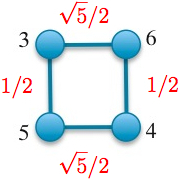}
\caption{(color online) Graph of a weighted square-cluster state corresponding to the squeezed operators in Eq.~(\ref{o1}). The vertices represent the intracavity
modes, the edges represent the coupling between these and fractions are the weightings.}
\label{figure1}
\end{figure}

In order to investigate the squeezed joint quadrature operators, and in turn determine whether the entangled output beams emerging from the cavity approximate a four mode cluster state, we begin by defining the quadrature field operators for each mode and arbitrary phase angle, $\theta$, as
\begin{equation}
\hat{X}_{i}=\hat{a}_{i}e^{-i\theta}+\hat{a}^{\dagger}_{i}e^{i\theta}, \hspace{0.2cm} \hat{Y}_{i}=\hat{X}_{i}(\theta+\pi/2).
\label{eqn1}
\end{equation}
The squeezed joint quadrature operators are then the eigenvectors of the system found by solving the Heisenberg equations of motion from the interaction Hamiltonian in the undepleted pump approximation. These are,
\begin{eqnarray}
O_{1}&=&(-c_{1}\hat{X}_{3}+c_{1}\hat{X}_{4}-\hat{X}_{5}+\hat{X}_{6})e^{-c_{2}r},\nonumber\\
O_{2}&=&(-c_{2}\hat{X}_{3}-c_{2}\hat{X}_{4}+\hat{X}_{5}+\hat{X}_{6})e^{-c_{1}r},\nonumber\\
O_{3}&=&(c_{1}\hat{Y}_{3}+c_{1}\hat{Y}_{4}+\hat{Y}_{5}+\hat{Y}_{6})e^{-c_{2}r},\nonumber\\
O_{4}&=&(c_{2}\hat{Y}_{3}-c_{2}\hat{Y}_{4}-\hat{Y}_{5}+\hat{Y}_{6})e^{-c_{1}r},
\label{o1}
\end{eqnarray}
where $c_{1}=(\sqrt{5}-1)/2$, $c_{2}=(\sqrt{5}+1)/2$ and $r$ is the squeezing parameter. The common eigenstate of these operators tends towards a quadripartite entangled cluster state when $r\rightarrow \infty$. The graph for the corresponding cluster state is depicted in Fig.~\ref{figure1}.

\subsection{The positive-$P$ representation equations}

We use the positive-$P$ representation~\cite{plusP} equations for the system described in Sec.~\ref{sys}. These are derived using the standard approach~\cite{GardinerQN} whereby the master equation is mapped onto a Fokker-Planck equation for the positive-$P$ function using a correspondence between two independent stochastic fields $\alpha_{i}$ and $\alpha^{+}_{i}$ and the mode operators $\hat{a}_{i}$ and $\hat{a}^{\dagger}_{i}$, respectively. The resulting equations can then be written as a set of $c$-number stochastic differential equations. This method allows us to calculate classical stochastic averages which are equivalent to quantum mechanical normally-ordered operator moments. 

The positive-$P$ equations for the high frequency fields are
\begin{eqnarray}
\frac{d\alpha_{1}}{dt}&=&\epsilon_{1}-\chi_{1}\alpha_{4}\alpha_{5}-\chi_{1}\alpha_{3}\alpha_{6}-\gamma_{1}\alpha_{1},\nonumber\\
\frac{d\alpha_{2}}{dt}&= & \epsilon_{2}-\chi_{2}\alpha_{5}\alpha_{6}-\gamma_{2}\alpha_{2},
\label{posp1}
\end{eqnarray}

\noindent and for the low frequency fields we find
\begin{eqnarray}
\frac{d\alpha_{3}}{dt}&=& \chi_{1}\alpha_{1}\alpha_{6}^{\tiny{+}}-\gamma_{3}\alpha_{3}+\sqrt{\frac{\chi_{1}\alpha_{1}}{2}}(\eta_{9}(t)+i\eta_{10}(t)),\nonumber\\
\frac{d\alpha_{4}}{dt}&=&\chi_{1}\alpha_{1}\alpha_{5}^{\tiny{+}}-\gamma_{4}\alpha_{4}+\sqrt{\frac{\chi_{1}\alpha_{1}}{2}}(\eta_{5}(t)+i\eta_{6}(t)),\nonumber\\
\frac{d\alpha_{5}}{dt}&=&\chi_{1}\alpha_{1}\alpha_{4}^{\tiny{+}}+\chi_{2}\alpha_{2}\alpha_{6}^{\tiny{+}}-\gamma_{5}\alpha_{5}+\sqrt{\frac{\chi_{1}\alpha_{1}}{2}}(\eta_{5}(t)-i\eta_{6}(t))\nonumber\\&+&\sqrt{\frac{\chi_{2}\alpha_{2}}{2}}(\eta_{1}(t)+i\eta_{2}(t)),\nonumber\\
\frac{d\alpha_{6}}{dt}&=& \chi_{1}\alpha_{1}\alpha_{3}^{\tiny{+}}+\chi_{2}\alpha_{2}\alpha_{5}^{\tiny{+}}-\gamma_{6}\alpha_{6}+\sqrt{\frac{\chi_{1}\alpha_{1}}{2}}(\eta_{9}(t)-i\eta_{10}(t))\nonumber\\&+&\sqrt{\frac{\chi_{2}\alpha_{2}}{2}}(\eta_{1}(t)-i\eta_{2}(t)).\nonumber\\
\label{posp}
\end{eqnarray}
Since the phase-space is doubled there are also the equations found by swapping $\alpha_{i}$ with $\alpha_{i}^{+}$ and $\eta_{i}(t)$ with $\eta_{i+2}(t)$. The $\eta_{i}(t)$ are real, independent, Gaussian noise terms with correlations $\overline{\eta_{i}(t)}=0$ and $\overline{\eta_{i}(t)\eta_{j}(t^{\prime})}=\delta_{ij}\delta(t-t^{\prime})$. 

\subsection{Linearized Fluctuation Analysis}

We now use Eqs.~(\ref{posp1}) and~(\ref{posp}) as the starting point for a linearized fluctuation analysis~\cite{Danbook}. Linearizing these around the classical steady-state solutions, we find the evolution equations for the fluctuations. Specifically, we neglect the noise terms in Eq.~(\ref{posp}) and set $\alpha_{i}=\bar{\alpha}_{i}+\delta\alpha_{i}$, where $\bar{\alpha}_{i}$ are the steady-state values and $\delta\alpha_{i}$ represent the fluctuations. To first order the equations of motion for the fluctuations, $\boldsymbol{\delta\alpha}=[\delta\alpha_{1}, \delta\alpha^{+}_{1},\delta\alpha_{2},\delta\alpha^{+}_{2},\dots,\delta\alpha_{6},\delta\alpha^{+}_{6}]^{T}$, can be written as~\cite{midgley}
\begin{equation}
d\boldsymbol{\delta\alpha}=-\boldsymbol{\bar{A}\delta\alpha} dt + \boldsymbol{\bar{B}} d\boldsymbol{W},
\end{equation}
where $\bf{\bar B}$ is the noise matrix of Eq.~(\ref{posp}) with the classical steady-state values inserted, $d\boldsymbol{W}$ is a vector of Wiener increments~\cite{GardinerQN} and $\boldsymbol{\bar A}$ is the drift matrix with the classical steady-state values inserted. 

Provided that the eigenvalues of the drift matrix $\boldsymbol{\bar A}$ have no negative real part, the system is well described as a multi-variable Ornstein-Uhlenbeck process~\cite{SMCrispin}, for which the intracavity spectral correlation matrix is found as
a matrix product which is then related to the measurable output fluctuation spectra by the standard input-output relations~\cite{collett}.

\section{Results and Discussion}

We now calculate the spectral variances of the squeezed joint operators and confirm that these approach zero over a range of phases and frequencies. We consider the symmetric system where the two pumping inputs, $\epsilon_{1,2}$, are equal, the two nonlinearities, $\chi_{1,2}$ are equal, and the cavity losses, $\gamma_{1,..,6}=\gamma$, are equal. As shown in Ref.~\cite{midgley}, the system has an oscillation threshold and for the parameters used here the critical pump amplitude is $\epsilon_{c}=61.8$. The parameters used in the following calculations are $\gamma=1$, $\chi=0.01$ and $\epsilon=0.94\epsilon_{c}$, giving a pump rate approximately 12\% below threshold in intensity. For these parameters, this was found to yield the best squeezing. 

\begin{figure}[tbhp]
\includegraphics[width=0.451\textwidth]{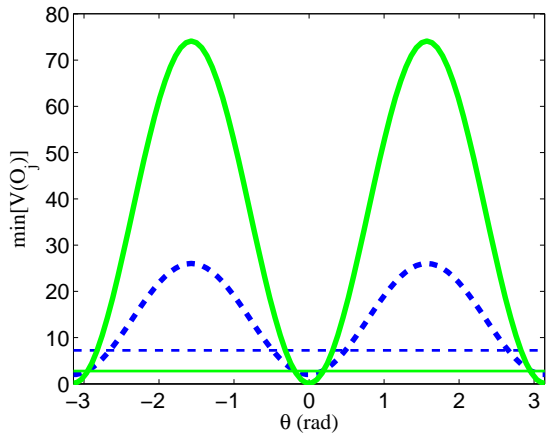}
\caption{(color online) Plot of the minimum variances of the squeezed joint operators as a function of phase at a fixed frequency of $\omega =$ 0.35$\gamma$.}
\label{figure2}
\end{figure}

In Fig.~\ref{figure2} we plot the minimum variance of the squeezed joint operators, $V(O_{j})$, as a function of the phase at a fixed frequency. We find the frequency at which the minimum variance is obtained and plot all four variances at this corresponding frequency. Specifically, Fig.~\ref{figure2} represents the maximum squeezing achieved by the operators $V(O_{1,3})$ but not $V(O_{2,4})$. The frequency that yields the maximum squeezing corresponds to $\omega =$ 0.35$\gamma$, corresponding to 30 MHz with a cavity loss rate 80 MHz. We find that $V(O_{1})=V(O_{3})$, as indicated by the (green) solid line, and $V(O_{2})=V(O_{4})$, as indicated by the (blue) dashed line. The corresponding flat lines indicate the coherent state level for the variances of the particular quadrature combinations, the dashed line being at 7.24 and the solid line at 2.76. This result shows squeezing well below the coherent state level and identifies a region where the squeezed joint operators approach zero for a range of phases centered around $\theta=0$ and multiples of $\pi$. This provides a theoretical quantum treatment of the experimental results obtained in Ref.~\cite{pysher} and provides further evidence that an ideal continuous variable four mode cluster state can be closely approximated by their single OPO scheme. 

\begin{figure}[tbhp]
\includegraphics[width=0.451\textwidth]{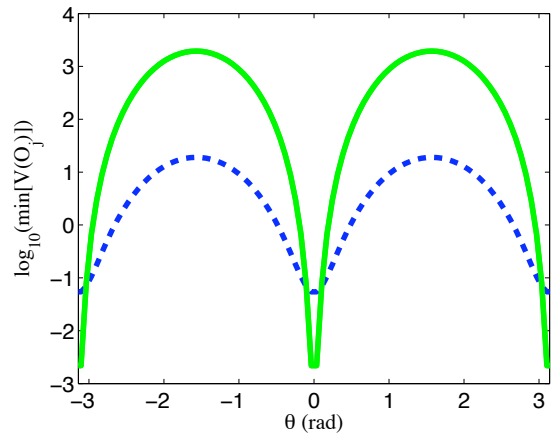}
\caption{(color online) Logarithmic plot of the minimum variances of the squeezed joint operators as a function of phase at a fixed frequency of $\omega =$ 0.35$\gamma$.}
\label{figure3}
\end{figure}

Figure~\ref{figure3} depicts the same quantity shown in Fig.~\ref{figure2} as a function of phase, but on a logarithmic plot. This plot compares well to the experimental results presented in Ref.~\cite{pysher} with the width of the squeezing region becoming smaller as the degree of squeezing increases. The squeezed joint operators are scaled relative to the relevant coherent state level of zero. 

\begin{figure}[tbhp]
\includegraphics[width=0.451\textwidth]{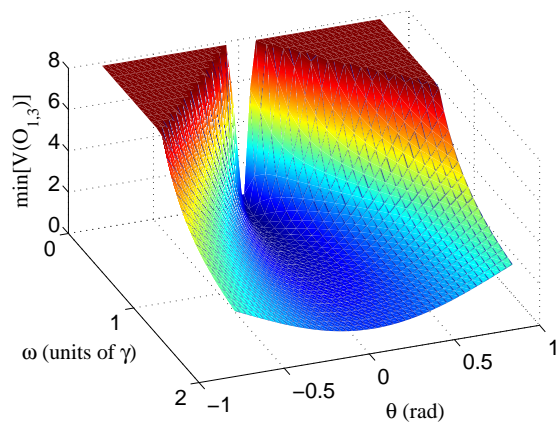}
\caption{(color online) Plot of the minimum variances of the squeezed joint operators, $V(O_{1})$ and $V(O_{3})$, as a function of phase and frequency.}
\label{figure5}
\end{figure}

\begin{figure}[tbhp]
\includegraphics[width=0.451\textwidth]{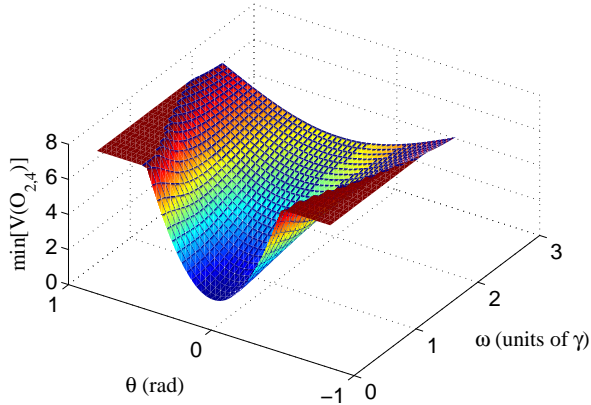}
\caption{(color online) Plot of the minimum variances of the squeezed joint operators, $V(O_{2})$ and $V(O_{4})$, as a function of phase and frequency.}
\label{figure6}
\end{figure}

We also investigate the behavior of the squeezed joint operators over a range of frequencies. Figures~\ref{figure5} and ~\ref{figure6} show plots of the minimum variances as a function of both frequency and phase. The range of parameters over which the entangled state can be considered as approximately a four mode cluster state can be seen in the smooth region surrounding $\theta=0$ and where $V(O_{j}) \rightarrow 0$. In both Fig.~\ref{figure5} and Fig.~\ref{figure6} the functions are cropped at $V(O_{j})=8$ to improve visualization. 

\begin{figure}[tbhp]
\includegraphics[width=0.451\textwidth]{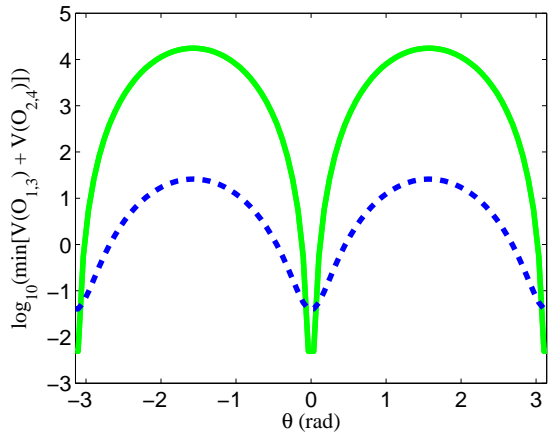}
\caption{(color online) Logarithmic plot of the minimum of the summed variances of both squeezed joint operator pairs as a function of phase at a fixed frequency of $\omega =$ 0.23$\gamma$.}
\label{figure7}
\end{figure}

Finally, we consider the regime of optimal squeezing for the joint operators. Figure~\ref{figure7} shows the minimum of the sum of the variances of both squeezed joint operator pairs as a function of phase on a logarithmic scale. Although this compromises the overall degree of squeezing obtained, this regime may be of use in experiments as the corresponding frequency of $\omega =$ 0.23$\gamma$ yields the best possible degree of squeezing for the combined quadratures.

\section{Conclusions}

In conclusion, we have demonstrated that the requirements of a CV four mode cluster state are closely satisfied in the single OPO scheme proposed in Ref.~\cite{men2} and experimentally realized in Ref.~\cite{pysher}. This is verified by confirming that the squeezed joint quadrature operators for the system approach zero for a range of phases and frequencies. These findings are in good agreement with the experimental results of Pysher~\etal~and extend current theoretical results~\cite{pysher} into the spectral domain. This work builds on our previous findings in Ref.~\cite{midgley} wherein CV quadripartite entanglement was demonstrated in the system and the squeezed joint quadrature operators were calculated in the temporal domain. Using our approach, it is a simple matter to extend the analysis to parameter regimes above the oscillation threshhold, which may have advantages when photon flux is important.\\
\section*{Acknowledgments}
MKO thanks the Australian Research Council for support under the Future Fellows scheme and SLWM would like to acknowledge the EPSRC.


\end{document}